\documentclass[referee]{raa}

\usepackage{graphicx,times}
\usepackage{natbib}
\usepackage{amssymb,amsmath}
\bibpunct{(}{)}{;}{a}{}{,}

\usepackage[pagebackref=true]{hyperref}

\begin{document}

   \title{Lunar nutation effect defines the sign of the Earth rotation rate for now, but this may change soon}

 \volnopage{ {\bf 20XX} Vol.\ {\bf X} No. {\bf XX}, 000--000}
   \setcounter{page}{1}

\author{Oleg Titov\inst{1,2,3} }  
\institute{Geoscience Australia, PO Box 378, Canberra 2601, Australia  \\
%\\ WWW home page:
%\texttt{http://users/\homedir iekeland/web/welcome.html}
\and
Phase$\&$Rate, Canberra 2600, Australia; 
\and
Shanghai Astronomical Observatory, 80 Nandan road, Shanghai 200030, China; \\
{\it oleg.titov@shao.ac.cn; andyrobinzon@hotmail.com}
}

%   \author{Oleg Titov  %% Put your Chinese name in "( )" if you like.
%   \inst{1}}
%   }
%% Here is an example of three authors come from different institutes.
%% For single author or all the authors from an institute, use "\inst{}" only

%   \institute{Geoscience Australia, PO Box 378, Canberra 2601, Australia; {\it andyrobinzon@hotmail.com}
%    \institute{Phase&Rate, Canberra 2615, Australia; {\it andyrobinzon@hotmail.com}
%    \institute{Shanghai Astronomical Observatory, China; {\it oleg.titov@shao.ac.cn}
%% Please give the E-mail address of the author, to whom future correspondence and
%% offprint requests will be sent.
%\\
%\vs \no
%   {\small Received 20XX Month Day; accepted 20XX Month Day}
%}

\abstract{The Earth slowly decelerates in its rotation due to the energy dissipation caused by the interaction with the Moon. This leads to the continuous increasing in the length of the mean solar 
day (aka, length-of-day, or, LOD) relatively to 86,400 solar seconds at an average secular rate of +1.8 ms per century. But, on a shorter time scale the process is uneven. A positive leap second is used to be introduced on regular basis to support a consistency between the astronomical and atomic timescales. However, nowadays the LOD is steadily sparking a discussion about the timescale maintenance, in particular, from fears that a negative leap second will have to be introduced for the first time in the foreseeable future. 
The aim is to show that the LOD is currently dominated by the 18.6 yr lunar nutation signal whereas the long-term trends are essential for extrapolation after 2030.
The LOD data since 1962 are used to estimate the long-term variations along the 18.6 yr and other harmonic signals in its spectrum. It is shown that the influence of the lunar nutation impact on the LOD variations was underestimated. At the moment, the LOD changes are completely determined by a signal with a period of 18.6 yrs. More detailed extrapolation reveals that LOD is likely to vary in a range between -1 ms to +1 ms until 2050 or may be longer.
}
   \authorrunning{O. Titov}            %author_head in even pages
   \titlerunning{Lunar nutation defines the Earth rotation}  

\maketitle

\section{Introduction}

\label{sec:intro}

Length-of-Day is gradually increasing due to the dissipation of energy in the Earth-Moon system due to tidal friction. Over a long-time interval, the secular slowdown in the Earth's rotation speed has an amplitude of +1.8 milliseconds per 100 years \citep{Stephenson2016}. In addition, the spectrum of the LOD contains various signals ranging from several minutes to several centuries. In particular, the spectrum contains signals with periods of 6 months, 1 yr, 9.3 yr and 18.6 yr (e.g. \citet{Lambeck_1973}, \cite{Gross1992}, \citet{Ray_2014}). The seasonal variations are caused by the global mass transport processes in the ocean and atmosphere; variations with periods of 9.3 and 18.6 yr are associated with the orbital motion of the Moon, which affects the LOD. 
Such a secular trend makes the timescales based on astronomical observations differ from the uniform atomic timescale based on the very stable hydrogen maser time and frequency standards. To ensure that the scales do not diverge, the International Earth Rotation and Reference Systems Service (IERS) introduces from time to time an extra second added to the atomic scale, also known as a leap second. This causes many technical problems, since our modern computer software systems do not support the introduction of a 61st second. As a result of software problems due to uneven adding of a leap second, a massive failure of computer systems can occur, which can lead to large financial losses \citep{Charles_2012, DeAngelis_2015}. For this reason, various ways to solve the 61st second problem have been discussed for a long time, however, despite all efforts, the community of scientists and engineers has not yet reached a consensus (\citet{Nelson_2001}, \cite{Levine_2023}).

Since about 2000, the slowdown of the Earth’s rotation has noticeably decreased to almost zero, therefore, only five leap seconds have been added in the 21st century (2005, 2009, 2012, 2015, 2017), and the last time a leap second was introduced was on 2017 January 1. Moreover, concerns have arisen that given the current trend in the Earth of rotation speed will continue, soon, it will be necessary to introduce an additional second with a minus sign – a negative leap second (e.g. \cite{Agnew_2024}). Since all of humanity has never dealt with a minute consisting of 59 seconds instead of 60, there is a good reason to believe that the number of computer failures with the introduction of a negative leap second will be much greater than with the introduction of familiar positive leap second. Therefore, if such a need arises, the consequences of the problem may cause issues with technical software on a planetary scale. 

The Earth's tides are not the only source of LOD variations. For example, \cite{Hopfner1998} presented estimates of the annual and semiannual harmonics in LOD series obtained by various authors (see also references therein). All of them are much larger than the theoretical values published in the IERS Conventions (2010), and for the annual harmonic the discrepancy is an order of magnitude \citep{Chen_2005}.  \cite{Hopfner1998} found that this discrepancy can be explained by seasonal transport of atmospheric masses. Although discrepancies in seasonal amplitudes have already been noted and explained, long-term changes in LOD still require study. \cite{Chao_2014} were the first to notice the abnormal amplitude of the lunar signal with a period 18.6 yr. \cite{Mouel_2019+} claimed a high amplitude of the 18.6 yr signal (0.65 ms) in time series of LOD from 1962 to 2018 while the theoretically predicted amplitudes do not exceed 0.2 ms. \cite{Zotov_2020+} detected a harmonic with a period of 18.9 years and an amplitude of about 1 ms based on the analysis of LOD residuals after subtracting the theoretical amplitude. Apparently, this is another indication of the discrepancy for the 18.6 yr amplitude, though, left unattended. As it will be shown in this manuscript, underestimating this signal can affect the LOD prediction over a short time interval.

Interannual variations of LOD with periods of 5-12 years were also reported by many authors, for example, the signals of 5.9 (or 6), 7.3, 7.8, 8.3, 8.6,  and 11.9 yr (e.g. \cite{Vondrak_1977, Silva_2012+, Holme_2013, Mouel_2019+, Duan_2020, Hsu_2021+, Ding_2021, Rosat_2023, Malkin_2024, Cazenave_2025+}).
Various physical processes are called that can cause signals with the specified periods - dynamical processes within the Earth's core, core-mantle interaction, variations of the Earth's geomagnetic field, solar activity, etc. Despite the general interest in these processes, it should be noted that the expected and measurable signal amplitudes in this frequency range are estimated at 0.1-0.15 ms, so it is unlikely that they could somehow "amplify" the tidal harmonic with a period of 18.6 yr. Therefore, in this article, we will limit ourselves to evaluating only two harmonics with periods of 5.9 and 11.9 yr, in order to assess the possible impact on the evaluation of tidal signals.

\section{Analysis of LOD data}
\label{sect:Obs}

\begin{figure}[ht]
    \centering
\includegraphics[width=0.5\linewidth]{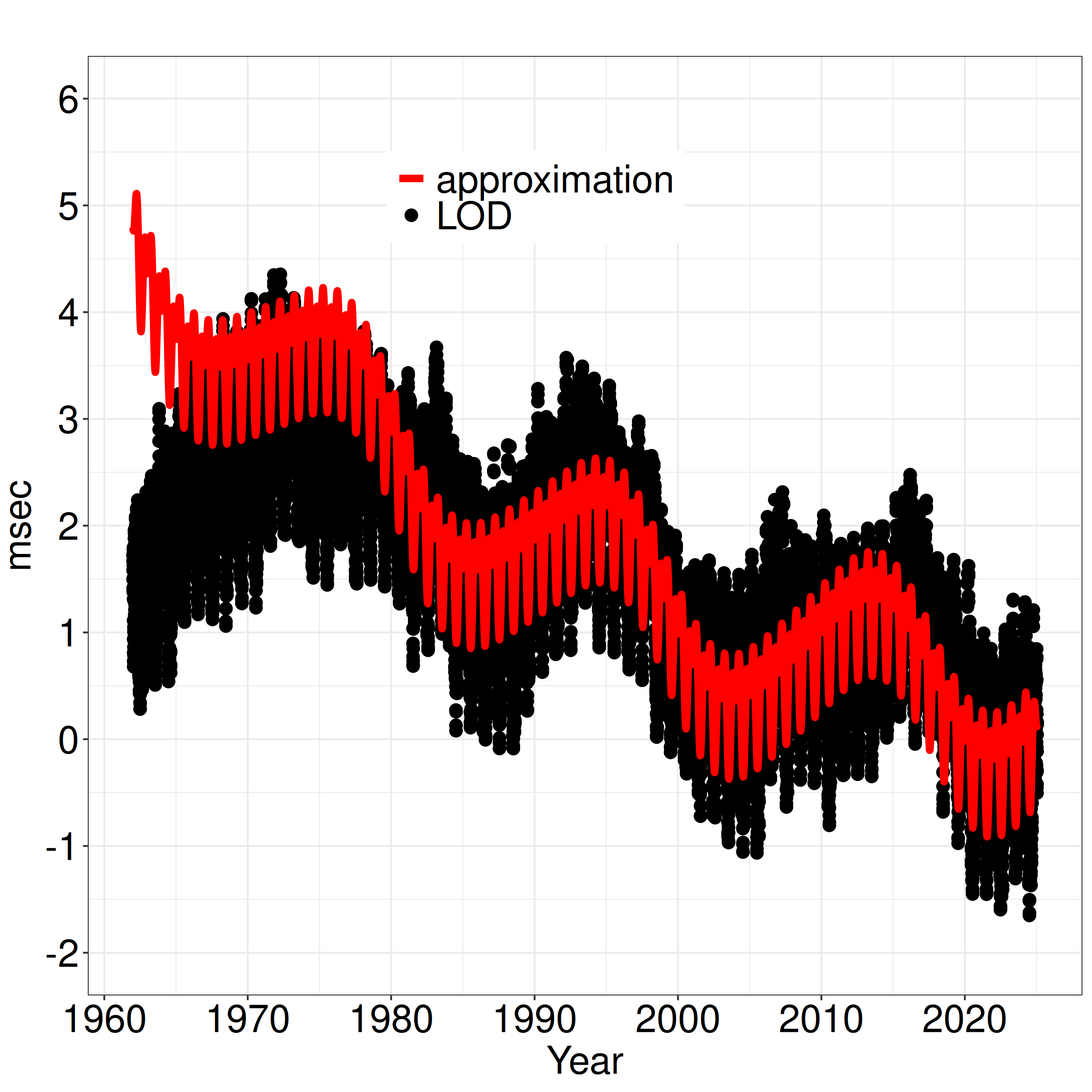}
    \caption{Fitting of the LOD data with the model including quadratic trend and four harmonic signals (Eq.~\ref{Eq1}) with amplitudes from Table~\ref{tab2}.  }
    \label{fig1}
\end{figure}

To assess the likelihood that a negative leap second will be introduced in the next 20 years, we will analyze the time series of LOD, starting from 1962, which is published by IERS(\footnote{\label{fnote1}\url{(https://www.iers.org)}}) \cite{Bizouard_2018+}. A time series containing more than 23,000 LOD daily points between 1962 and 2025 is shown on Figure~\ref{fig1}. 
To approximate this time series using the least squares method, we take a simple model containing a linear (or quadratic, or cubic) time trend and four harmonic signals with periods of 0.5, 1, 9.3 and 18.6 yr, which were previously noted by various authors (\citet{Ray_2014}, \citet{Agnew_2024}). Equation~(\ref{Eq1}) displays the parametrical model for fitting using the least squares method

\begin{equation}
\begin{aligned}
  \frac{\Delta LOD}{LOD} = \sum^{k}_{i=0} a_{i} \Delta t^{i} + 
 \sum^{4}_{j=1} (A^{c}_{j}\cos\frac{2\pi\Delta t}{P_{j}} + 
 A^{s}_{j}\sin\frac{2\pi\Delta t}{P_{j}})   \\
	\label{Eq1} % Use a logical label
\end{aligned}
\end{equation}

where 
%$\frac{\Delta LOD}{LOD}$
LOD is in ms/day 
(note as LOD measured in ms throughout the paper!),  $\Delta t = T - T_{0}$, $T$ is the observation date in years, and $T_{0}=1993.525$ is a zero epoch for the parametrization of the time trends in Equation~(\ref{Eq1}). 
The index $k$ in Equation~(\ref{Eq1}) is a power of the polynomial used, and the index $i$ is running from 0 to $k$ in the first sum. $P_{j}$ are the period of the four harmonic signals $({j=1,2,3,4})$ and $A^{c}_{j}, A^{s}_{j}$ are two amplitudes of the signals to be estimated.

\begin{figure}[ht]
    \centering
\includegraphics[width=0.5\linewidth]{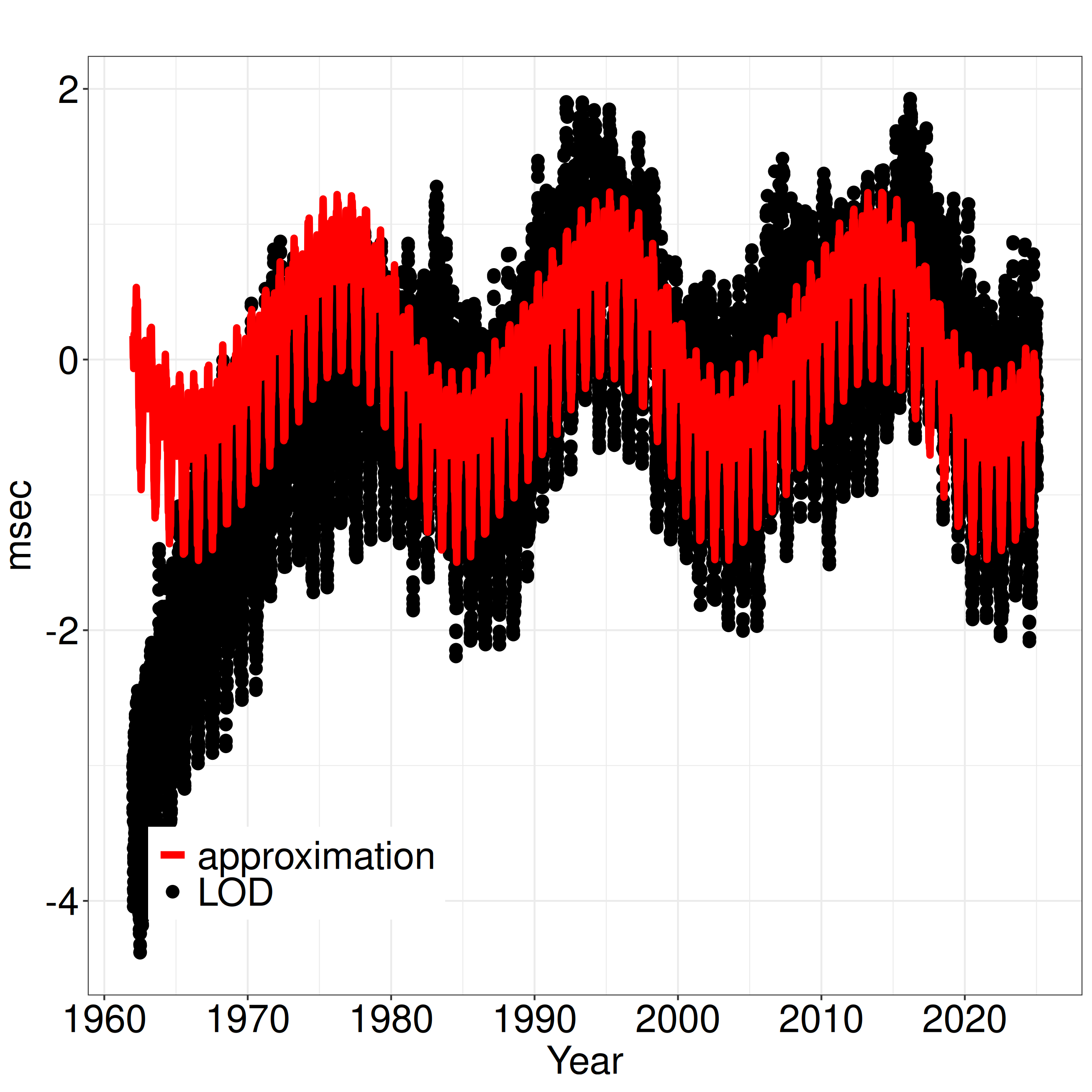}
    \caption{Fitting of the LOD data with the model including four harmonic signals only (Eq.~\ref{Eq1}) with amplitudes from Table~\ref{tab2}.  }
    \label{fig2}
\end{figure}

\begin{figure}[ht]
    \centering
\includegraphics[width=0.5\linewidth]{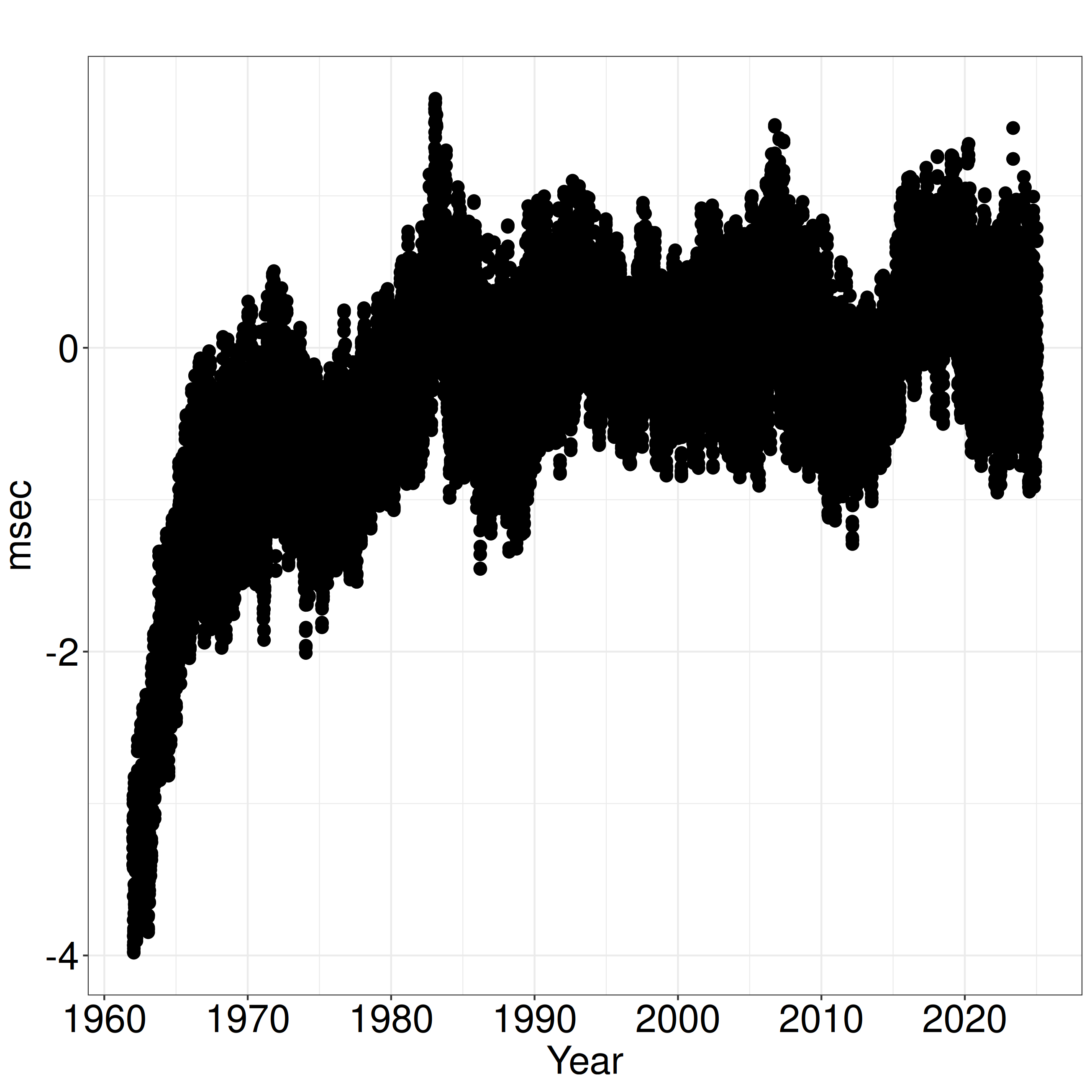}
    \caption{Post-fit residuals after removing of all systematic fitted with  (Eq.~\ref{Eq1}). }
    \label{fig3}
\end{figure}

\begin{table}[ht]%
	\centering
	% Captions go above tables
	\caption{Estimates of the parameters for the largest polynomial terms in the Eq~(\ref{Eq1}) for LOD }
	\label{tab1} % give each table a logical label name	
	\begin{tabular}{lcccl} 		\\
		\hline \
		i & & $a_{i}$ & & units \\
    		 & k=1  & k=2 & k=3 &  \\
		\hline
       \\
		0 & 1484 $\pm$ 5  & 1577 $\pm$ 5  &  1591  $\pm$ 5 & $\mu$s\\
		1 & -38.8 $\pm$ 0.3   & -68.3 $\pm$ 0.8  & -67.0 $\pm$ 0.8   &  $\mu$s/year\\
        2 &   & 1.01 $\pm$ 0.03 & 0.62 $\pm$ 0.07 & $\mu$s/year$^{2}$   \\
        3 &  &  & 0.012 $\pm$ 0.002  & $\mu$s/year$^{3}$  \\
		\hline
	\end{tabular}
\end{table}

\section{Results}

\subsection{Least squares fitting}

The least squares estimate for trends and harmonic signals obtained with Eqaution~(\ref{Eq1}) are in Tables \ref{tab1} and \ref{tab2}. While the estimated coefficients for the time trend are roughly consistent with the results published by other authors, the amplitude estimates for the four tidal harmonics are much higher than previously published theoretical values in Table \ref{tab3} (\citet{iers10}, \citet{Ray_2014}). A study of available sources showed that the previously published amplitudes were only based on theoretical data, and quite a long time ago. This anomalously strong observed amplitude of the 18.6 yr harmonic was already noted previously, but without any consequences, e.g. \citep{Chao_2014, Bizouard_2018+, Mouel_2019+}. For example, \cite{Ray_2014} showed a theoretical curve, which obviously does not match observations, and ascertained that in the observed LOD series, the 18.6 yr line is obscured by decadal variability from nontidal processes. However, as it could be seen on Figures ~\ref{fig1} and  ~\ref{fig2}, it looks reasonable simply to assume that the observed 18.6 yr signal is much stronger than the theory prescribes, instead of referring to some weird nontidal signals.  It is unclear whether the discrepancy found is a flaw in the theory, or there is additional stronger theoretical effect with an 18.6 yr period in the LOD time series. However, regardless of its interpretation, this is a key factor for understanding of the current LOD variations. 

\vspace{0.4cm}

\begin{table}[ht]
	\centering
	% Captions go above tables
	\caption{Estimates of the parameters for the model of four tidal harmonics in Eq~(\ref{Eq1}) for LOD for the case $k$ = 2 (quadratic trend) and two sets of observations (1962-2025) and (1984--2025). Reference epoch $T_{0} = 1993.524$.}	
	\label{tab2} 
    % give each table a logical label name
	\begin{tabular}{lrrcrrc} 		
		\hline
		Period & $A^{c}_{i}$ & $A^{s}_{i}$ & Amplitude & $A^{c}_{i}$ & $A^{s}_{i}$ & Amplitude\\
		year & $\mu$s & $\mu$s & $\mu$s & $\mu$s & $\mu$s & $\mu$s \\
          & & 1962 --  2025 & & & 1984 -- 2025 & \\
		\hline
        \\
	    0.5 & -322 $\pm$ 4 & -79 $\pm$ 4  & 332 & -322 $\pm$ 5 & -81 $\pm$ 5  & 332 
        \\
        1.0 & -354 $\pm$ 4 & -132 $\pm$ 4  & 378 & -353 $\pm$ 5 & -133 $\pm$ 5  & 378
        \\
        9.3 & -53 $\pm$ 4 & 67 $\pm$ 4  &  85 &  -56 $\pm$ 5 & 65 $\pm$ 5  & 86   
        \\
        18.6 & 614 $\pm$ 4 & 231 $\pm$ 5  & 656  & 611 $\pm$ 4 & 232 $\pm$ 5  & 654\\
		\hline
\end{tabular}
\end{table}

\vspace{0.1cm}

\begin{table}[ht]
	\centering
\caption{Amplitudes of the four periodic terms: column 2 – estimates from the IERS online tool; column 3 – theoretical values from the IERS Conventions 2010 \citet{iers10}; column 4 - theoretical values from \citep{Ray_2014} , column 5 - theoretical values from \citep{Defraigne_1999}}
	\label{tab3} 
    % give each table a logical label name
	\begin{tabular}{lcccr} 		\\
		\hline
		Period &  & Amplitude $\mu$s & \\
		year &  &  &  &  \\
		\hline
        \
	    0.5 & 388 & 171 & 173 & 140  \\
        1.0 & 325 & 27 & 28 & 22 \\
        9.3 & 86 & 1  & 2 &  - \\
        18.6 & 655 & 149 & 160 & 126 \\
		\hline
\end{tabular}
\end{table}

\vspace{0.1cm}

For additional verification, the same analysis was performed using the interactive system on the IERS website (\footnote{\label{fnote2}\url{(https://www.iers.org/IERS/EN/Home/home_node.html)}}). A series of the same values of LOD was approximated by the same model from a quadratic trend and the four systematic harmonics. The fitted results are also shown in Table \ref{tab3}. These estimates are consistent with our results and completely contradict the predictions from theoretical works and IERS Conventions (\cite{iers10}). 

Figures~\ref{fig1} and ~\ref{fig2} show that the 18.6 yr component dominates the series, and, as the estimated amplitude of the 18.6 yr harmonic exceeds 0.6 ms, the total swing between maximum and minimum of the corresponding LOD systematic fraction reaches 1.3 ms. Therefore, most of the leap seconds introduced for the last three decades were entered on the timeline when the 18.6 yr signal was near its maximum values. For example, an “active” decade in 1990s, when eight additional seconds were added during that 10 yr interval, is followed by a long 6 yr break. The last three leap seconds were introduced during the last maximum of the 18.6 yr signal at the mid of 2010s (2012, 2015, 2017). In accordance with the empirical analysis, now we are near the minimum values of the red curve, but the moment of the minimum has already been passed, and the values of LOD are in the ascending phase of the 18.6 yr cycle.
 As now we are on the ascending phase of the signal, LOD is moving towards the positive leap second margin that could be seen in Figure~\ref{fig1}.

It should be noted that the formal accuracy of the early LOD data is much worse than that of the modern data. For example, the error of 1962 data is 1.4 ms, which is roughly 50 times larger than modern estimates in 2021-2025 (0.02-0.03 ms). The formal uncertainties of the LOD up to 1984 are quite significant (about 0.15 ms during 1983), so their contribution to least squares estimates is negligible, since the corresponding weights of early data used for fitting are 2500 (in 1962) and 25 (in 1983) times smaller than the weights of observations in 2021--2025. As a result, large residual deviations in 1962--1983 (Figure~\ref{fig3}) may create a misleading visual impression that the obtained results are unreliable. The right side of Table \ref{tab2} shows the amplitudes of the four tidal harmonics estimated along with the quadratic trend if a limited set of data between 1984 and 2025 is used. These results are almost identical to those in the left side of Table \ref{tab2}, even though the used time series is about 40 percent shorter.

\vspace{0.3cm}

\subsection{Extrapolation}

\begin{figure}[ht]
    \centering
\includegraphics[width=0.5\linewidth]{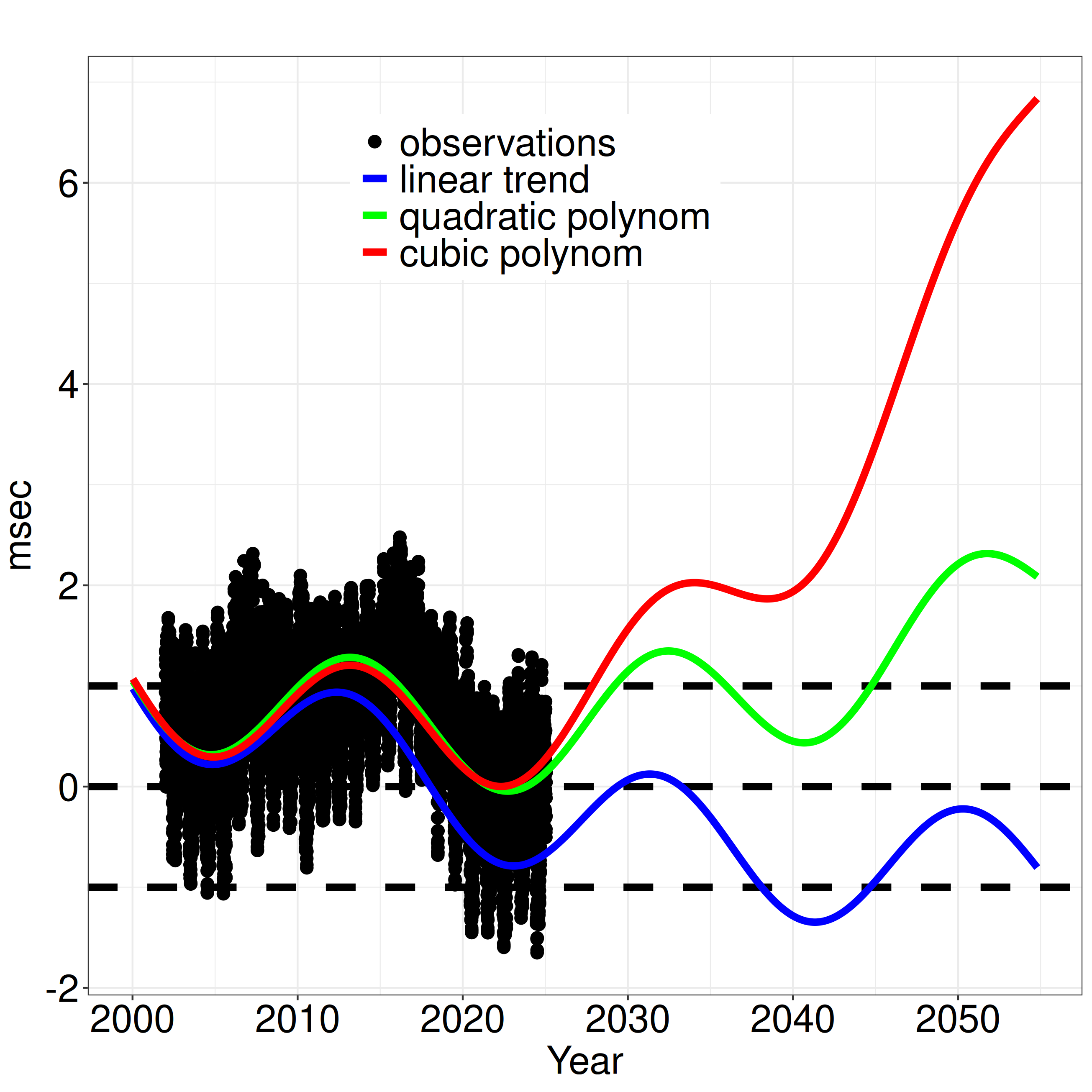}
        \caption{Extrapolation of the LOD observations to 2050 using three long-term trends. Horizontal dash lines show the level -1 ms, 0 ms, and +1 ms. }
    \label{fig4}
\end{figure}

One can argue that the empirical approximation very well represents variations in the LOD over almost the entire interval of observations, except for a short interval in the early 1960s which could be attributed to insufficient accuracy of the early observations compared to the modern data. Therefore, we will try to extrapolate the LOD behavior until 2050 based on a model that includes a long-term trend and the 18.6 yr signal from Equation~(\ref{Eq1}) with the amplitude from the first column of Table~\ref{tab2}. The results of the extrapolation using the three long-term trends are presented in Figure~\ref{fig4} suggesting three scenarios until 2050.

\vspace{0.2cm}
Option 1 - linear trend
\vspace{0.3cm}

This option apparently discussed in the paper \citep{Agnew_2024} is based on the long-term influence of a stable linear trend, and similar conclusions follow. The LOD variations are in a comfortable corridor between -1 ms and +1 ms for quite a long time. There is a risk of introducing a negative second after 2040, although this may be ignored until next minimum epoch of the lunar cycle (near 2060), as it relies on the stability of the predicted negative linear trend. 

\vspace{0.4cm}
Option 2  - quadratic trend
\vspace{0.3cm}

 The quadratic term presents graphically an ascending parabola, so we are currently at the lowest point of the green curve in Figure \ref{fig4}. The LOD values will gradually increase after 2022, and the negative second will never need to be entered. The Earth's rotation will slow down further, and the next positive second will have to be added around 2030. Further, this process will gradually increase, though, it would be a period of relief until 2045. 

\vspace{0.4cm}
Option 3 - cubic trend
\vspace{0.3cm}

 A worse version of the previous scenario, and similar to those proposed by \cite{Malkin_2024}. A combination of quadratic and cubic polynomials will lead to serious consequences shown in red in Figure \ref{fig4}. Slowing down of the Earth's rotation will go so fast that the first positive second will be added before 2030. In the first half of the 2030s, the yearly gain for the constant value of LOD = +0.2 ms would result in the integrated value of extra LOD of 365~$\times$~0.2 ms = 0.73 second. Therefore, the Earth deceleration rate will proceed as in the 1990s, when a new leap second was added almost every year. After 2035, this process will take on an avalanche-like character. An additional second will have to be added even more frequently, if the approach to the formation of the UT1 scale does not change. 
 
\vspace{0.2 cm}

A fourth option (“the heavenly calm”) which presents a mix of options 1 and 2 may be considered, if nature comes to the aid of humanity once again. In this case, a negative linear trend and a positive quadratic trend will compensate each other until 2050, approximately. Then we will have enough time to find an alternative technical solution that will satisfy everyone's needs.

\section{Discussion}

It is necessary to realize that extrapolation always depends on the way the data was approximated for the previous (and very limited) time interval, and the chosen method of approximation by polynomials is not always a good choice. However, one could believe that the actual LOD variation behavior will be within the visual space between the red and blue lines on Figure~\ref{fig4}. In this situation, all that remains is to closely monitor observations over the next two years. If, by the beginning of 2028, the LOD values, cleared of seasonal components, will be around 1 ms, then the situation is developing according to Option 3, and we hope this will never happen. If the LOD values are equal or less than 0.5 ms, then it will be necessary to think about how to introduce the very first negative leap second by 2040. All intermediate values in a range between 0.5 and 1 ms in early 2028 will project to LOD values between 0 and 6 ms by 2050.

\begin{figure}[ht]
    \centering
\includegraphics[width=0.5\linewidth]{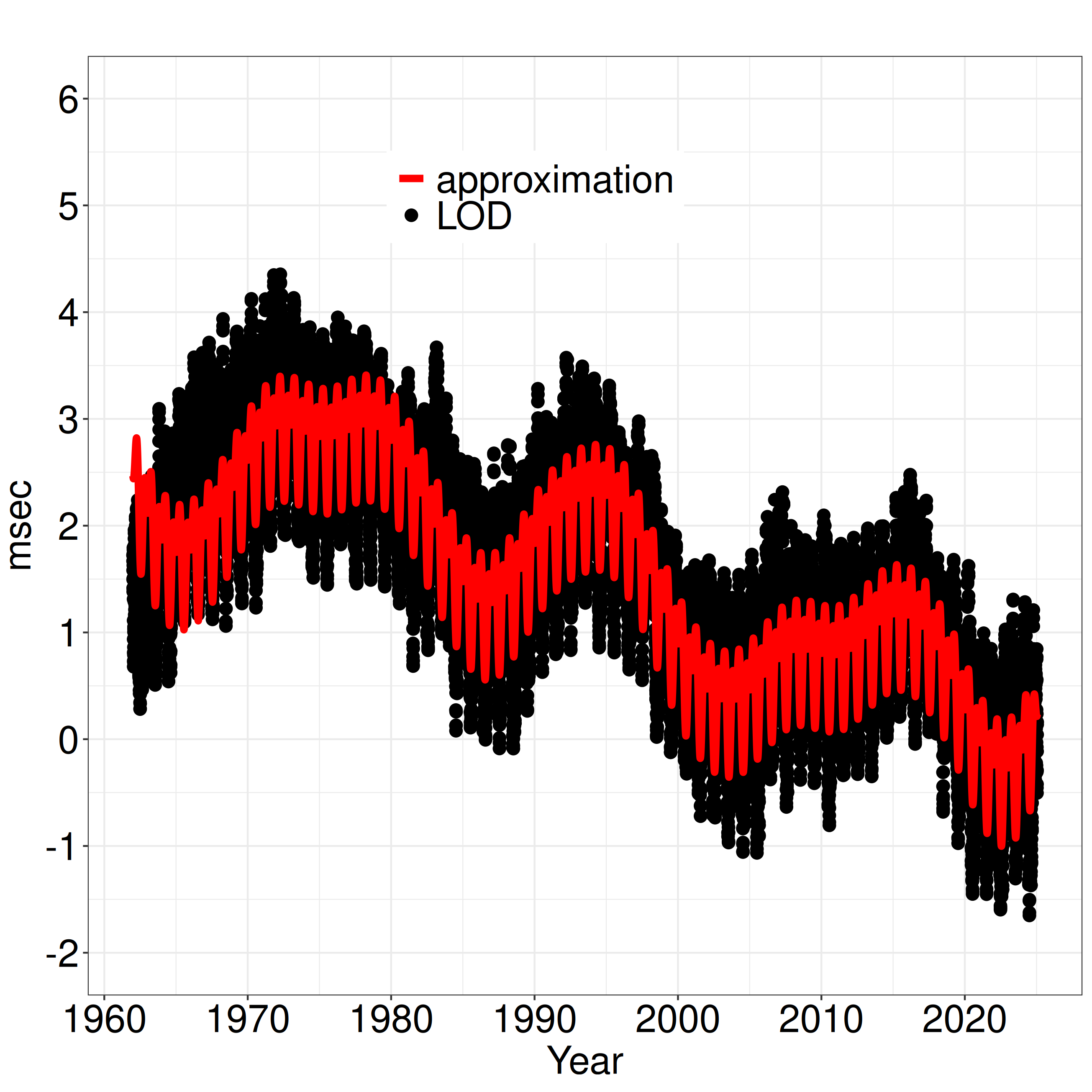}
    \caption{Fitting of the LOD data with the model including quadratic trend and four harmonic signals (Eq.~\ref{Eq1}) with amplitudes from Table~\ref{tab4}.  }
    \label{fig5}
\end{figure}

\begin{figure}[ht]
    \centering
\includegraphics[width=0.5\linewidth]{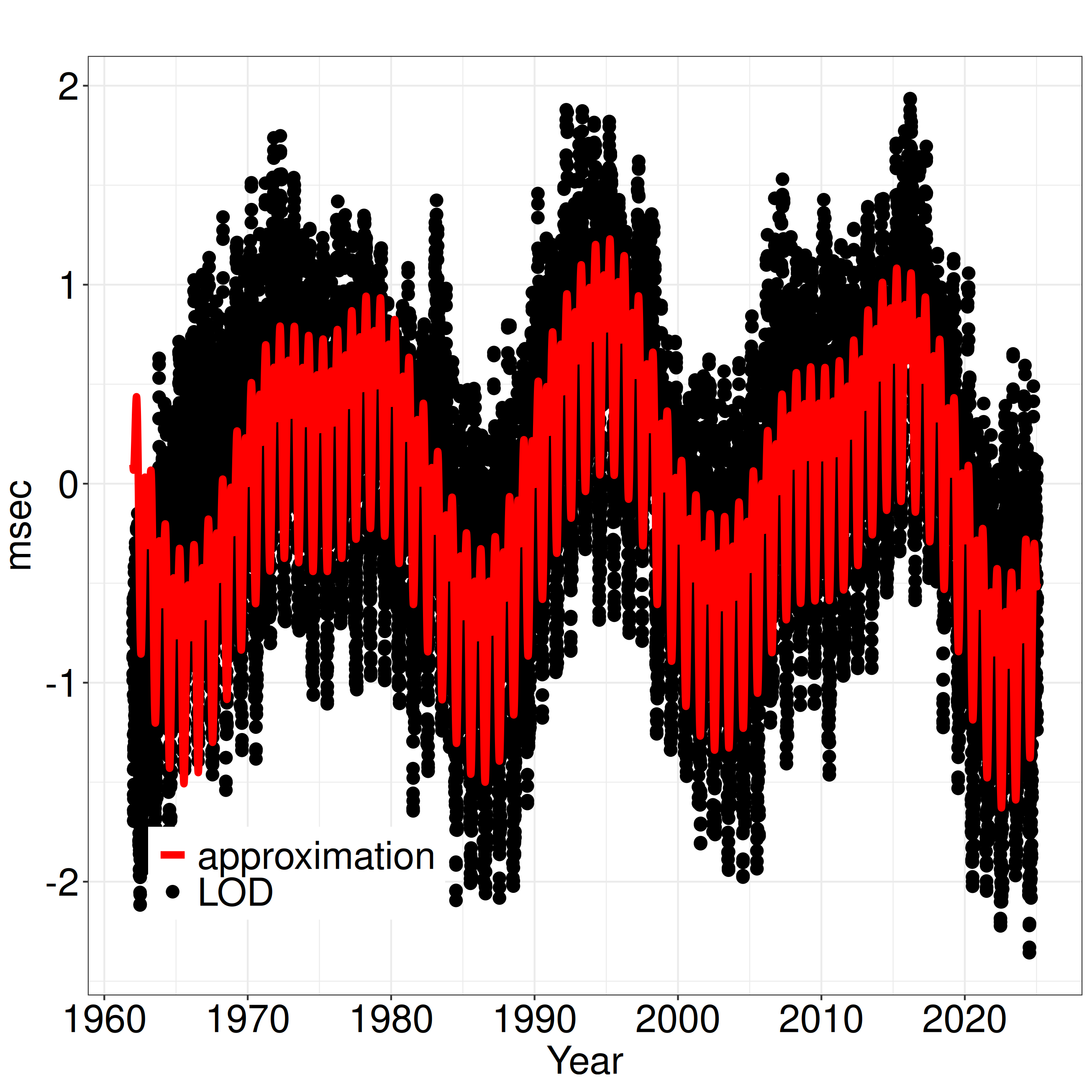}
    \caption{Fitting of the LOD data with the model including four harmonic signals only (Eq.~\ref{Eq1}) with amplitudes from Table~\ref{tab4}.  }
    \label{fig6}
\end{figure}

\begin{figure}[ht]
    \centering
\includegraphics[width=0.5\linewidth]{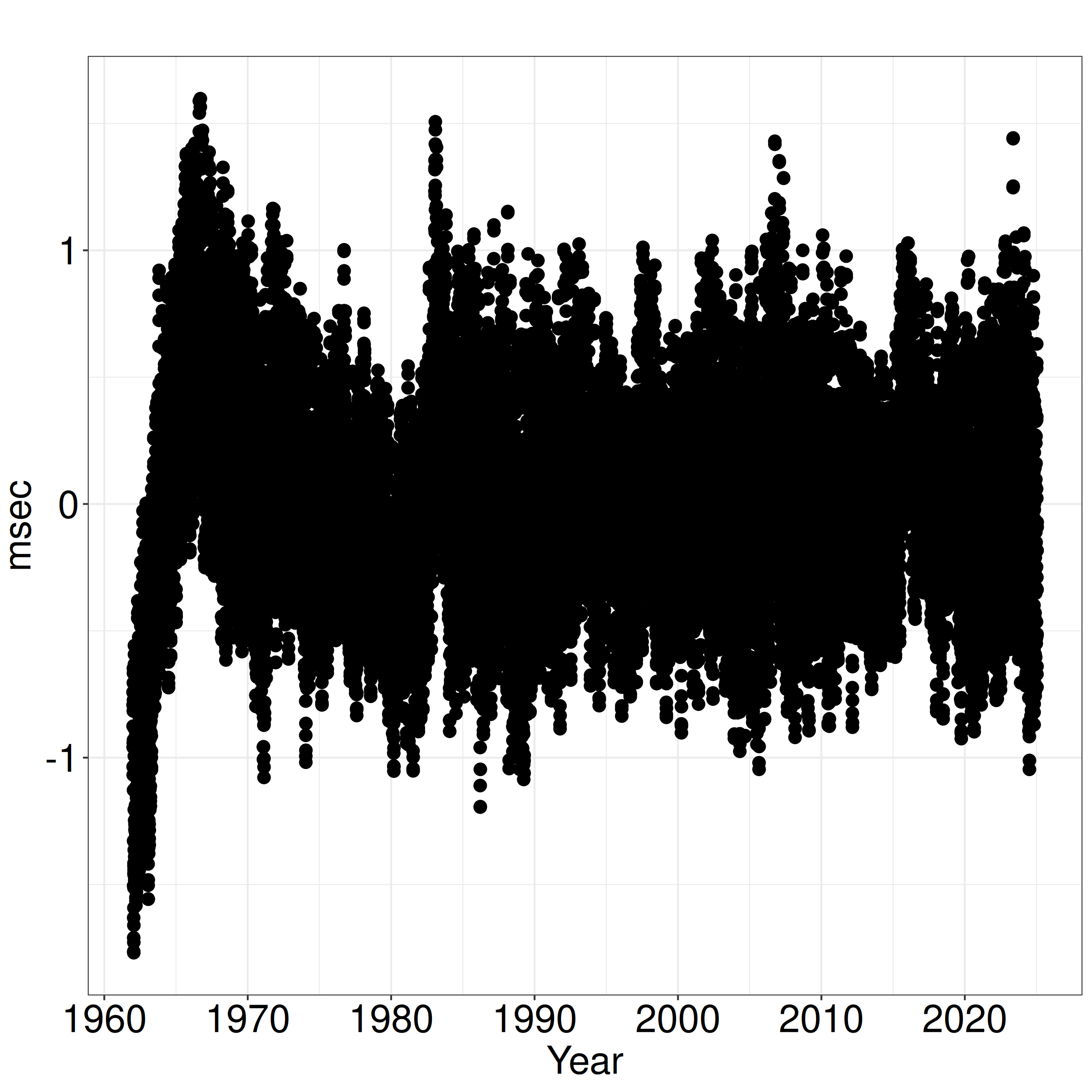}
    \caption{Post-fit residuals after removing of all systematic fitted with six harmonics from Table~\ref{tab4}. }
    \label{fig7}
\end{figure}

\vspace{0.4cm}

\begin{table}[ht]
	\centering
	% Captions go above tables
	\caption{Estimates of the parameters for the model of six harmonics and quadratic and cubic trends.}	
	\label{tab4} 
    % give each table a logical label name
	\begin{tabular}{lccccccc} 		
		\hline
		Period &  & Quadratic trend & &  & Cubic trend & \\
 & $A^{c}_{i}$ & $A^{s}_{i}$ &  Amplitude & $A^{c}_{i}$ & $A^{s}_{i}$ & Amplitude\\
        
year & $\mu$s & $\mu$s & $\mu$s & $\mu$s & $\mu$s & $\mu$s\\
		\hline
        \\
	    0.5 & -325 $\pm$ 4 & -78 $\pm$ 4  & 334 & -327 $\pm$ 4 & -77 $\pm$ 4  & 336
        \\
        1.0 & -349 $\pm$ 4 & -133 $\pm$ 4  & 374 & -349 $\pm$ 4 & -135 $\pm$ 4  & 374 
        \\
    5.9 & -24 $\pm$ 4 & 21 $\pm$ 4  & 32 & -31 $\pm$ 4 & 11 $\pm$ 4  & 33
        \\
        9.3 & -95 $\pm$ 4 & 2 $\pm$ 4  & 95 & -98 $\pm$ 4 & 21 $\pm$ 4  & 101   
        \\
    11.9 & 195 $\pm$ 4 &  50 $\pm$ 4  & 201 & 222 $\pm$ 4 &  50 $\pm$ 4  & 228
        \\
        18.6 & 566 $\pm$ 4 & 273 $\pm$ 5  & 629 & 565 $\pm$ 4 & -268 $\pm$ 5  & 625 \\
		\hline
\end{tabular}
\end{table}

As, the inter-decade variations of LOD  were reported by many authors, additional study was performed to assess a potential impact of other harmonics on the tidal amplitudes. The spectral analysis of the LOD variations also shows a presence of the two signals with period 5.9 and 11.9 yr after removal of the trend and four tidal harmonics from the original LOD time series. An additional analysis was conducted with six harmonics instead of four.  Table \ref{tab4} shows the amplitudes of the six harmonics fitted with the same method (Equation~\ref{Eq1}) with a quadratic or a cubic trend applied. Comparison of Tables \ref{tab2} and \ref{tab4} shows that the amplitudes of the four tidal harmonics hardly change. Only the main signal with a period of 18.6 years in Table \ref{tab4} becomes slightly smaller due to coupling with the newly introduced 11.9-year harmonic. 

The $\chi^2$ statistical parameter, which is equal to $\chi = 11.74$ for the four harmonic approximation, reduces to $\chi = 11.19$ and $\chi = 11.14$ after 11.9 yr and 5.9 yr signals subsequently added to the model, while the correlation coefficients between the four harmonics (5.9, 9.3, 11.9, 18.6 yr) do not exceed 0.3 in absolute value. Therefore, one could conclude that adding new harmonics to Equation~(\ref{Eq1}) does not significantly affect the amplitudes of the four original tidal harmonics. By visual impression the approximation of the original LOD time series in Figures~\ref{fig5}  and \ref{fig6} is better also at the early epochs, and the post-fit residuals (Figure~\ref{fig7}) show smaller systematic variations after removal of two additional harmonics. So, the harmonics with periods 5.9 and 11.9 yr are seem to be real and their interpretation is still an open problem.

\vspace{0.5cm}

\section{Conclusion}

Over the past 60 years or so, LOD changes have been linked to global processes in the atmosphere, oceans, and Earth's core. These processes interrupt the secular positive trend, so the LOD values by 2020 decreased to zero. Here we show that the main role in determining the sign of the leap second is now played by the signal of the Moon's orbital nodal precession with a period of 18.6 yr and peak-to-peak range about 1.3 ms. The tiny long-term negative linear trend from Table \ref{tab2} was exaggerated by that 18.6 yr signal on its descending phase during the time interval between 2015 and 2022. This probably caused some prediction of necessity of a quick introduction of a negative leap second in the mid of 2020s followed by a potential threat of software malfunction. However this 18.6 yr signal has been on the ascending phase since 2022, therefore, there is no an immediate threat of a negative leap second until the next well-predicted minimum that is expected near 2040, if the linear trend is on the same slope. A more reliable forecast could be made in beginning of 2028 when the behavior of the LOD time series becomes clearer.

The reasons for the large discrepancies between theory and observations for the four periodical signals in Table \ref{tab2} are not discussed here. Once the semiannual and annual signals may be attributed to the impact of the seasonal atmosphere mass transfer, it is necessary to find out why theoretical models of 9.3 and 18.6 yr tidal variations differ so greatly from observational data. However, since the amplitudes of all four harmonics predicted by the theory are much fewer than observational ones, then, apparently, the theoretical scatter of points during extrapolation will be much less than what is shown on Figure~\ref{fig4}. Thus, such a noticeable discrepancy between theory and observations does not affect the main conclusion of this study in any way.

\normalem
\begin{acknowledgements}

This paper is published with the permission of the CEO, Geoscience Australia. The authors acknowledge use of VLBI data obtained within the framework of the International VLBI Service for Geodesy and Astrometry (IVS). Length of Day (LOD) results published by the International Earth Rotation and Reference System Service (IERSS). It was supported by the Astrometric Reference Frame project, Grant No. JZZX-0102, Shanghai Astronomical Observatory. 
I am grateful to Dr. Christian Bizouard from Paris Observatory for his comments and interesting discussion.

\end{acknowledgements}

\bibliographystyle{raa}
\bibliography{ms2025-0450}

\end{document}